# Mermin's Inequalities of Multiple qubits with Orthogonal Measurements on IBM Q 53-qubit system

Wei-Jia Huang[1], Wei-Chen Chien[2], Chien-Hung Cho[1], Che-Chun Huang[1], Tsung-Wei Huang[3], Ching-Ray Chang[2,4]

[1]*Department of Physics, National Taiwan University, Taipei, Taiwan*

[2] *Graduate Institute of Applied Physics, National Taiwan University, Taipei, Taiwan*

[3] *Department of information and computer Engineering, Chung Yuan Christian University, Taiwan*

[4]*Department of Business Administration, Chung Yuan Christian University, Taiwan*

**Entanglement properties of IBM Q 53-qubit quantum computer are carefully examined with the noisy intermediate-scale quantum (NISQ) technology. We study GHZ-like states with multiple qubits (N=2 to N=7) on IBM Rochester and compare their maximal violation values of Mermin's polynomials with analytic results. A rule of N-qubits orthogonal measurements is taken to further justify the entanglement less than maximal values of local realism (LR). The orthogonality of measurements is another reliable criterion for entanglement except the maximal values of LR. Our results indicate that the entanglement of IBM 53 qubits is reasonably good when N ≤ 4 while for the longer entangle chains the entanglement is only valid for some special connectivity.**

## Introduction

Entanglement is a very unique feature of the quantum sciences and cannot be observed in the classical world. Coherence is another general property of waves and describes the correlation between the constituent parts. Both entanglement and coherence are the fundamental elements of the quantum world. The Bell inequality[1] defines a boundary between quantum non-locality and local realism (LR) for double quantum states and Mermin[2] extended the results into an entanglement of a higher number of quantum states. Mermin's inequalities[2] is an excellent test for the entanglement property of a state for checking whether it violates LR. Quantum correlation of measurements for an entangled system always exist even the sub-systems of entangled states are physically separated far away. This entanglement provides a possibility to study the information of all entangled states simultaneously and gives a superior advantage of quantum computation than classical computation[3].

There are various examples about the second quantum revolution with quantum computers[4]. The violation of Bell inequalities has been verified in atomic physics experiments[5,6]. Daniel Alsina *et al.*[7] has reported Mermin's inequalities[2] on IBM 5-qubit quantum computer. The search algorithm for highly entangled states of multiple qubits have also been discussed[8,9]. Recently, IBM 53-qubit quantum computer (IBM Rochester)[10] is online for all kinds of experiments. However, the coherence time of their qubits can only support the realization of a noise intermediated state quantum (NISQ)[11] environment and very few things can be done in reality, so we should try to reduce the environmental noise[12]. Alternative way is to emulating quantum computers on classical platforms[13]. Until now only some simple examples were demonstrated on the IBM quantum computer system[14-16]. However, in near-term, algorithms like the quantum adiabatic optimization algorithms[17] and the variational quantum eigensolvers[18] have some chance of demonstrating quantum advantage[19] for a N-qubits system. In addition, there are also other applications like hash preimage attacks[20] and modeling viral diffusion[21]. Google Sycamore achieved quantum supremacy[22] and took only 200 seconds to perform certain problems that the Summit supercomputer need 10,000 years to compute.

The full understanding of the entanglement properties of a large number of qubits within state-of-art quantum computers, e.g., IBM Rochester, becomes critical for the real applications. Previous studies between quantum non-locality and classical LR are mainly focused on the static and theoretical situations for a small number of qubits, in particular for pure quantum states as Bell and Greenberger-Horne-Zeilinger (GHZ)-type states[23]. Only recently had experimental data been reported but even that were still not a fully dynamic measurement due to the limitation of coherence time and the small number of mixed entangled states[7,24,25]. Therefore, a full understanding of the entanglements and correlations between large number of qubits are important for a universal quantum computer. Here we propose to use orthogonal measurements of Mermin's inequalities[2] to study the influence of phase angles and the correlations of orthogonal measurements in GHZ-like states[7]. Also a longer chains of qubits are executed on IBM Q 53-qubit quantum computer[10] with the nearest-neighboring connectivity. The entanglement and correlations of Mermin's polynomials from 2 to 7 qubits are carefully examined. The orthogonal measurements give a consistent result with the analytic results under NISQ.

**Theory**

There are some indicators to evaluate the performance of a quantum computer, for example, negativity[26], quantum volume[27], or Mermin's inequalities[2]. However, the major difference between classical physics and quantum physics mainly is still mainly the entanglement property. A 2-qubits system can use either Bell's inequality[1] or Mermin's inequalities[2] to check whether the system violates LR. For a larger number of qubits, it needs to study Mermin's polynominals[28] to understand the entanglement of N-qubit and their quantum subsystems. The goal of experimental test on IBM Rochester is to characterize the entanglement behavior of N-qubits, and GHZ-like state[7] can assure the Mermin's polynomials will be maximum. Thus, here a GHZ-like state with appropriate initial phases[29] is used to probe the entanglement of N-qubits on the IBM 53-qubits machine.

For N qubits, the recursive relation of Mermin's polynomials are

$$\begin{cases} M_n = M_{n-1}(a_n + a'_n) + M'_{n-1}(a_n - a'_n) \\ M'_n = M_{n-1}(a'_n - a_n) + M'_{n-1}(a_n + a'_n) \end{cases} \quad (1)$$

where $M_1 = a_1 = X$, $M'_1 = a'_1 = Y$, $a_n = X$, $a'_n = Y$, and integer $n \geq 2$.

Operator $X$ and operator $Y$ represent Pauli-$X$ gate and Pauli-$Y$ gate respectively.

The GHZ-like state[7] is

$$|\Psi_n\rangle = \frac{1}{\sqrt{2}}(|00\ldots0\rangle + e^{i\varphi}|11\ldots1\rangle) \quad (2)$$

$\varphi$ is the phase angle and the expectation value of Mermin's polynomials are

$$\langle M_n \rangle = \begin{cases} 2^{n-\frac{1}{2}}\cos\left(\varphi - \frac{n-1}{4}\pi\right), & n = even \\ 2^{n-1}\cos\left(\varphi - \frac{n-1}{4}\pi\right), & n = odd \end{cases} \quad (3)$$

$$\langle M'_n \rangle = \begin{cases} 2^{n-\frac{1}{2}}\sin\left(\varphi - \frac{n-1}{4}\pi\right), & n = even \\ 2^{n-1}\sin\left(\varphi - \frac{n-1}{4}\pi\right), & n = odd \end{cases} \quad (4)$$

Please see *supplement A* for the details of the derivations for $\langle M_n \rangle$ and $\langle M'_n \rangle$.

The upper limit of LR is [2]

$$\langle M_n \rangle_{LR} \leq \begin{cases} 2^{\frac{n}{2}}, & n = even \\ 2^{\frac{n-1}{2}}, & n = odd \end{cases} \quad (5)$$

For brevity, we normalize all $\langle M_n \rangle$ and $\langle M'_n \rangle$ values by $\langle M_n \rangle_{LR}$ and thus

$$\widetilde{M}_n = \frac{\langle M_n \rangle}{\langle M_n \rangle_{LR}} = 2^{\frac{n-1}{2}} \cos\left(\varphi - \frac{n-1}{4}\pi\right), \quad (6)$$

$$\widetilde{M'}_n = \frac{\langle M'_n \rangle}{\langle M_n \rangle_{LR}} = 2^{\frac{n-1}{2}} \sin\left(\varphi - \frac{n-1}{4}\pi\right), \quad (7)$$

The maximal values of $\widetilde{M}_n$ and $\widetilde{M'}_n$ with the appropriate phase angle $\varphi_{max}$ are listed in Table I.

| number of qubits: $n=$ | 2 | 3 | 4 | 5 | 6 | 7 |
|---|---|---|---|---|---|---|
| $\varphi_{max}$ | $\frac{1}{4}\pi$ | $\frac{1}{2}\pi$ | $\frac{3}{4}\pi$ | $\pi$ | $\frac{5}{4}\pi$ | $\frac{3}{2}\pi$ |
| $e^{i\varphi_{max}}$ | $\frac{1+i}{\sqrt{2}}$ | $i$ | $\frac{-1+i}{\sqrt{2}}$ | $-1$ | $\frac{-1-i}{\sqrt{2}}$ | $-i$ |
| $\frac{\text{maximal value}}{\langle M_n \rangle_{LR}} = 2^{\frac{n-1}{2}}$ | $\sqrt{2}$ | $2$ | $2\sqrt{2}$ | $4$ | $4\sqrt{2}$ | $8$ |

Table I: Maximal values of $\langle M_n \rangle$ and $\langle M'_n \rangle$ are normalized by $\langle M_n \rangle_{LR}$. Here $\varphi_{max}$ is phase angle which $\langle M_n \rangle$ has its maximal value.

From Eqs. (6) and (7), it is important to note there is a sum rule for $\widetilde{M}_n^2 + \widetilde{M'}_n^2 = 2^{n-1}$ for quantum entanglement among N-qubits system. For $\widetilde{M}_n$ and $\widetilde{M'}_n$, the orthogonal measurements within an entangled system are strongly correlated, this unique quantum sum rule is very different from the independent and isotropy nature of a classical system.

**Method**

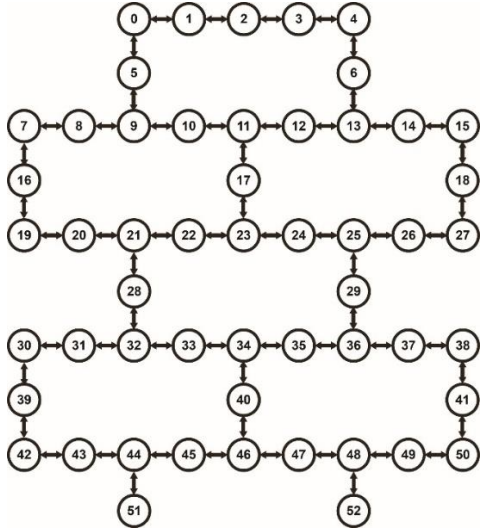

Fig. 1: The connectivity of qubits within hexagonal lattice structure of IBM Q 53-qubits quantum computer (IBM Rochester).

To study the entanglement of N-qubits, the adjacent qubits within the hexagonal structure of IBM Rochester are studied and all combinatory of N-qubits which are analyzed in this article are listed in *supplement B*. The connecting structure of IBM Rochester is shown in Fig. 1.

A single qubit initialized with $|0\rangle$ state becomes the superposition state $\frac{1}{\sqrt{2}}(|0\rangle + |1\rangle)$ with an operation by the *H* gate[30]. When a CNOT gate is applied, those other qubits entangled with the superposition state are shown in Fig. 2. Then a *U1* ($\varphi$) gate operates on the qubit to form the complete quantum circuit for the N-qubits GHZ-like state. For example, in the case of a 5-qubits GHZ-like state, we set $|GHZ_5\rangle = \frac{1}{\sqrt{2}}(|00000\rangle + e^{i\varphi}|11111\rangle)$ as in Fig. 2(a).

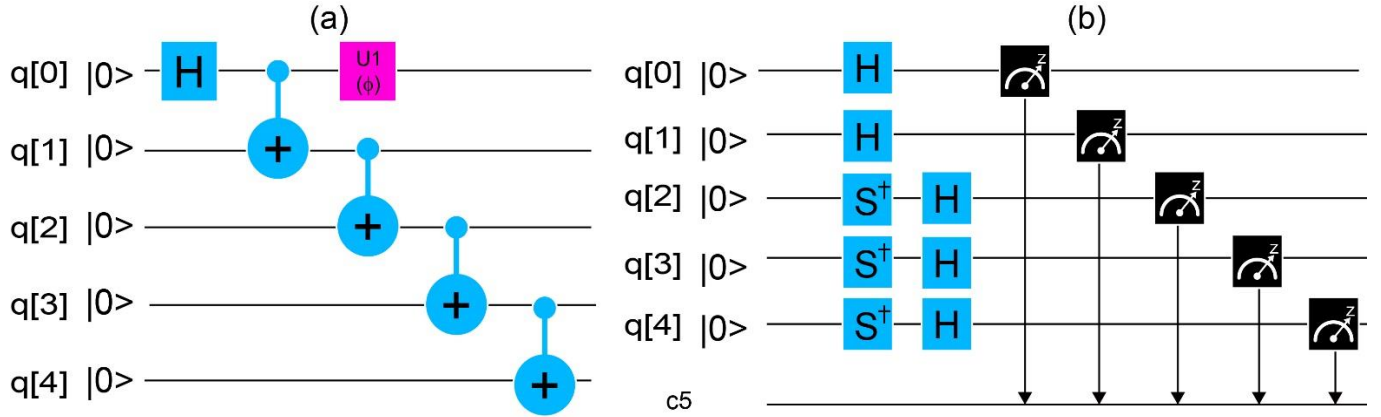

Fig. 2: (a) The quantum circuit for 5-qubits GHZ-like state, $|GHZ_5\rangle = \frac{1}{\sqrt{2}}(|00000\rangle - |11111\rangle)$. *H* gate is Hadamard gate, $U1\ (\varphi)$ is a gate to rotate along with the Z-axis with phase φ and *CNOT* gate entangles two qubits together. (b) The quantum circuit of *XXYYY measurement for 5-qubits*.

To measure the output of the quantum circuit, *H* gate needs to be included before a measurement along $\sigma_x$. To measure $\sigma_y$, both $S^+$ and *H* gates are needed before measurement. The operation of $S^+$ gate equates to a rotation of the state around the Z-axis with the angle $-\pi/2$, and this changes $\sigma_x$ to $\sigma_y$. For example, *XXYYY* measurement for the 5-qubits is shown in Fig. 2(b). Here *X* and *Y* gates are Pauli-*X* and Pauli-*Y* gate respectively.

**Experiment results**

Combinatory qubits with different qubit-connectivity within the hexagonal structure of the IBM Rochester are measured as Mermin's polynomials up to a 7-qubits connectivity. As shown in Fig. 3, the results from (a) to (f) are the measurement values of the Mermin's polynomials for different combinations of 2 to 7 entangled qubits on the IBM Rochester.

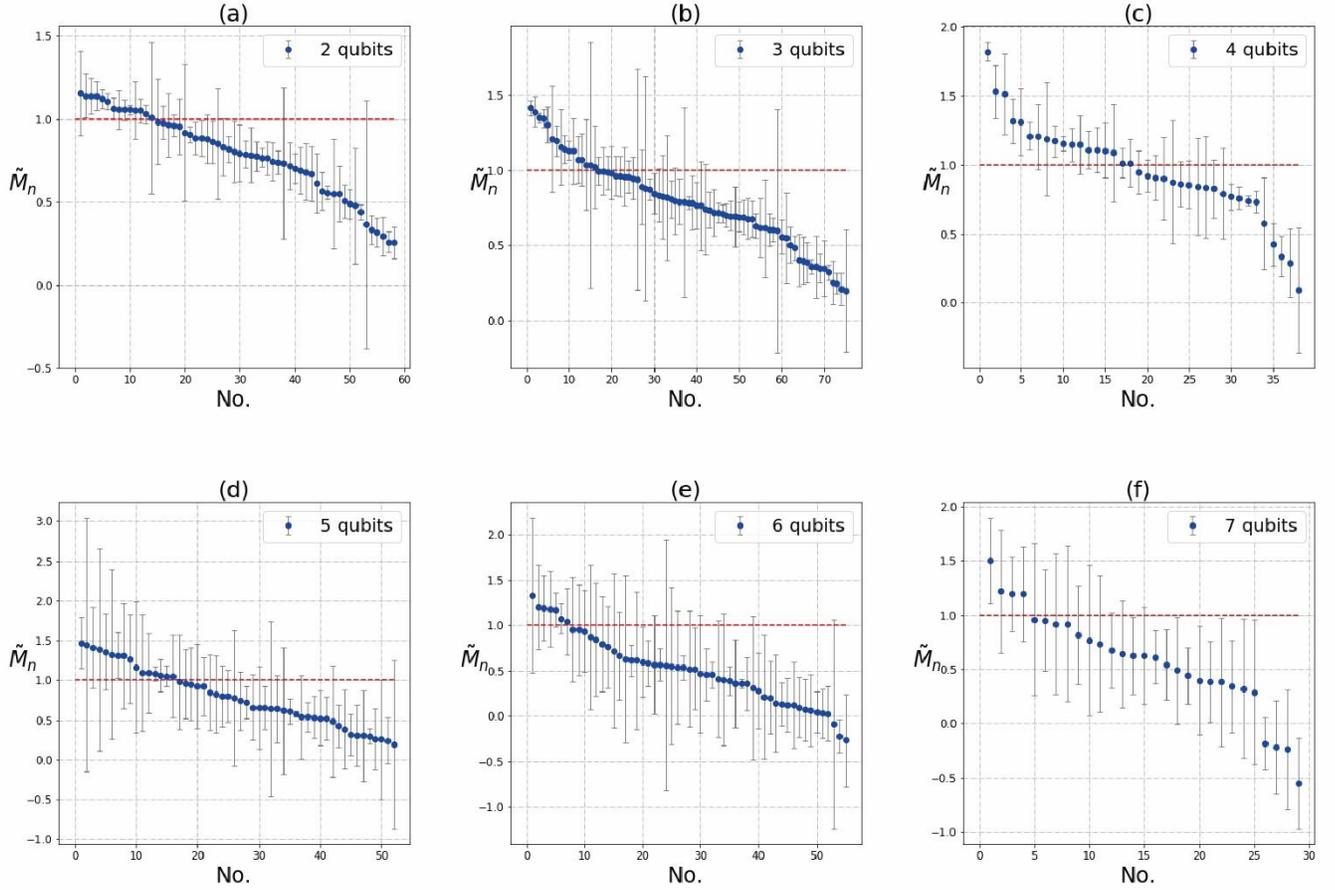

Fig. 3: The N-qubits GHZ-like states measurement of maximal expectation value of Mermin's polynomials. (a) 2 qubits (b) 3 qubits (c) 4 qubits (d) 5 qubits (e) 6 qubits (f) 7 qubits. The horizontal axis is the sequence number of connectivity for the N-qubits (listed in *supplement B*) chosen from the IBM Rochester hexagonal structure in Fig. 1. The vertical axis is $\widetilde{M_n}$, the measurement values of the Mermin's polynomials normalized by the maximal LR values. The measurement values are arranged in a descending order from left to right and the real connectivity of the qubits under measurement are listed in *Supplement B*. Experimental shots are 1024 and each data are the average of five measurements, and both the average values and standard variations (the bar length of the data point) are plotted.

The red line is the N-qubits LR of the Mermin's inequalities and the measurement values are arranged in a descending order from left to right. We choose the initial phase angle as shown in Table I to have a maximal value of $\langle M_n \rangle$ for the N-qubits. For the 2-qubits and the 3-qubits, all possible combinatory qubits within the IBM Rochester have been measured. We select all possible connectivity of 58 combinations for the 2-qubits and 75 combinations for the 3-qubits on the IBM Rochester. The outcomes indeed depend on the environments and the connectivity of the IBM Rochester. Each data point within Fig. 3 is the average of five measurements and experimental shots are assigned as 1024 for each measurement. The entanglement of qubits is apparently justified when the measurement value is larger than LR (red dotted lines with unity value). However, not all measurement results violate the LR of Mermin's inequalities (as shown in Fig. 3). Quantum system with small number of qubits show much better entangled results than systems with large number of qubits as evident in our criteria test based on the LR limit.

Only the connectivity of 3-qubits which violates the LR is chosen to connect to an extra qubit to form the new combinatory of 4-qubits for experiments on the IBM Rochester (*see supplement B*). Higher N-qubits connectivity is chosen in similar ways up to 7-qubits. All data in the Fig. 3 are $\widetilde{M_n}$ which is normalized with the maximal LR values of Mermin's inequalities. From Fig. 3 only a few combinations of the Mermin's polynomials exceed the predictions of LR value. However, from an analysis of the 2-qubits with an entangled parameter θ (*Supplement C*), the noise environment can reduce the effective entanglement strength due to the fluctuation of amplitudes of the superposition states. Thus the LR value cannot be the only criterion for the justification of entanglement. The statistical error for higher number of qubits are much larger than that for the small number of qubits as shown in the increasing standard deviations of Fig. 3 and *Supplement B*. The large standard deviations also imply significant influences from NISQ for larger qubit number. Therefore, studies beyond the criterion of the maximal values of Mermin's polynomials need to be taken.

To further understand if entanglement of the N-qubits exists in the IBM Rochester even with their measurement values less than the LR values, orthogonal measurements between $\langle M_n \rangle$ and $\langle M'_n \rangle$ are plotted. To study robustness of entanglement for the IBM Rochester 53-qubit system, we propose an orthogonal measurement (*supplement A*) to study the entanglement strength for systems with LR below the maximal value. A method reported by Pan *et. al*[31] with Mermin's inequalities for multiple number of qubits are extended to check the entanglement with a chosen initial phase angle and its associated $\langle M_n \rangle$ and $\langle M'_n \rangle$. From Eq. (3), the phase angle of the N-qubits with a maximal value of $\langle M_n \rangle$ is used for the orthogonal measurements between $\langle M_n \rangle$ and $\langle M'_n \rangle$. For brevity, in Fig. 4 we normalize all expectation values with the maximal value of LR. For entanglement N-qubits, the measurement values of $\langle M_n \rangle$ and $\langle M'_n \rangle$ should be orthogonal ( see Eqs. (6) and (7)). Intuitively for a classical system, the results of $\langle M_n \rangle$ and $\langle M'_n \rangle$ should be independent and isotropic. It can be observed that from Fig. 4(a), the orthogonal measurements of N-qubits for $N \leq 4$, the data points are more or less along the $\langle M_n \rangle$ axis even though the experimental values do not exceed LR. However, for a classical results, $\langle M_n \rangle$ and $\langle M'_n \rangle$ should be independent measurements and the outcome cannot be along the $\langle M_n \rangle$ axis with $\langle M'_n \rangle \sim 0$. The data for 2, 3, and 4 qubits, almost all combination of qubits are close to the $\langle M_n \rangle$ axis, even though data fall within the LR limit. This indicates the orthogonality of $\langle M_n \rangle$ and $\langle M'_n \rangle$ are reasonably good from the orthogonal criterion point of view. Therefore, the N-qubits for $N \leq 4$ can be concluded as an entangle system but with a reduced entangled strength arising due to the variation of amplitudes of the superposition states of NISQ (see *supplement C*).

For Fig. 4(b), measurements for the 5, 6 and 7 qubits obviously deviate from the $\langle M_n \rangle$ axis and quite symmetrically distribute along the diagonal direction of LR square. Only a handful of data outside the classical RL limit is observed. From the measurements of larger number of qubits, the probability of orthogonal measurements is almost the same even though the initial phase angle of the maximal value along the $\langle M_n \rangle$ axis is applied. For N-qubits with $N \geq 5$, $\langle M_n \rangle$ and $\langle M'_n \rangle$ are quite independent and isotropic as classical expectations and the data scatter along the diagonal direction. This indicates that for the $N \geq 5$ qubits system, entanglement within the NISQ is largely destroyed. Therefore, we can conclude entanglement for large number of qubits in the IBM Rochester are not robust and might easily drop. However, the IBM Rochester did show excellent entanglement properties for the N≤4 systems, but the entangled strength depends on the environment noise and varies with the connectivity of the qubits. For larger number of qubits, i.e., N≥5, the entanglement of N-qubits is not stable and the errors and standard variation become large probably due to the limited coherence time of NISQ. It can be concluded that for small number of qubits, i.e., 2, 3, and 4, entanglement is fine; but for large number of qubits, the results of entanglement are not robust.

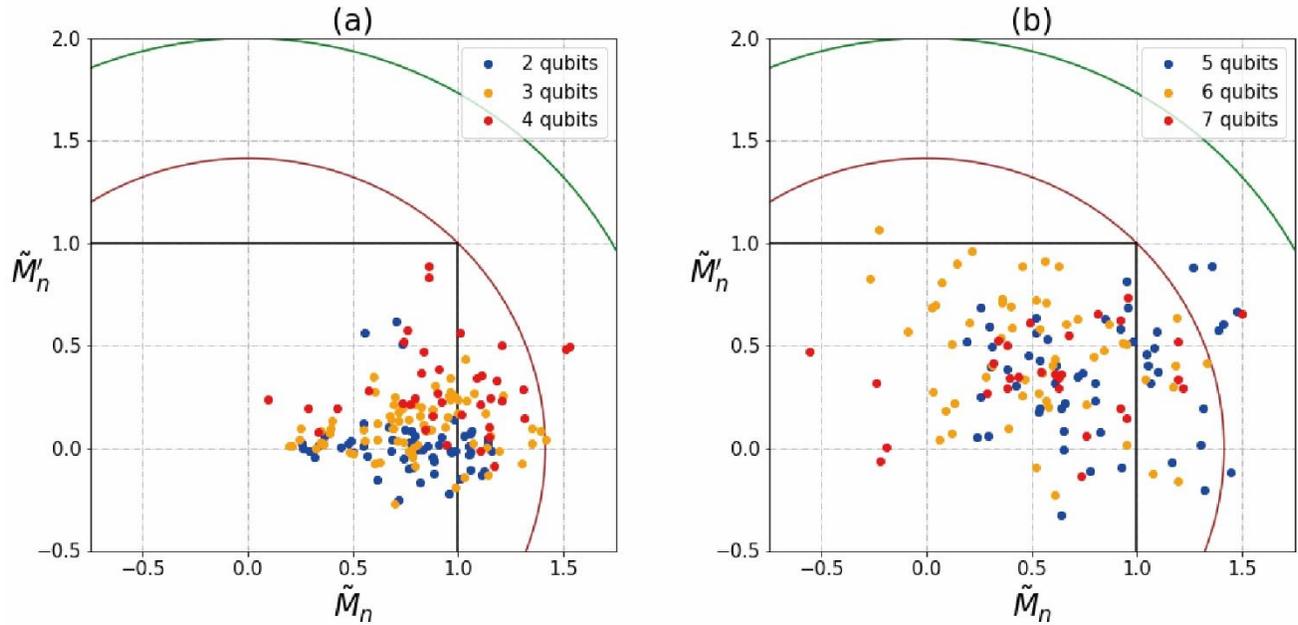

Fig. 4: The relationship between orthogonal measurements of $\widetilde{M_n}$ and $\widetilde{M'_n}$ is shown. The square corner is the upper limit for LR, the inner circle is the theoretical value of the 2-qubit subsystem entangled for a N-qubit system and outer circle is the for the 3-qubit subsystem of a N-qubit system. Experimental shots are 1024, and each data are the average of five measurements. (a) The experimental results of 2, 3, 4 qubits from different connectivity of IBM 53-qubit Rochester system. (b) The experimental results of 5, 6, 7

qubits from different connectivity of IBM 53-qubit Rochester system.

**Discussion and conclusion**

In summary an easy orthogonal measurement test between $\langle M_n \rangle$ and $\langle M'_n \rangle$ is proposed to examine the entanglement among the N-qubits when the maximal values of the Mermin's polynomials are within LR. We have used GHZ-like states to study the entangled pairs on the IBM Rochester 53 quits quantum computer and most of the pairs we have studied either had the orthogonal properties between $\langle M_n \rangle$ and $\langle M'_n \rangle$ or violated LR of Mermin's inequalities Therefore, two qubits are indeed entangled under NISQ and the IBM Rochester quantum computer performs reasonably well with 2-qubit entanglement. The entanglement is fine even up to the 4-qubits but with a reduced entangled strength from noise environment. While for the $N \geq 5$ qubits, only very few cases can be observed as entangled systems and also the entanglement strength fluctuates heavily for different combination of connectivity. Therefore, we can conclude that the IBM Rochester 53 qubits system is good for problems with usage of small entangled qubits under the NISQ environments. For large number of qubits, only some particular qubit connectivity can still remain robust in noise environments.

**DATA AVAILABILITY**

All data supporting the findings of this study are available from the authors upon request.

**AUTHOR CONTRIBUTIONS**

W.-J. H. and C.-R. C. contributed primarily to the design and the implementation of the research, the analysis of the results, and the writing of the paper. W.-C. C. C.-H. C and C.-C. H were involved in some parts of the design and implementation.

**Conflict of Interest Statement**

The authors declare no competing interests.

**Correspondence** and requests for materials should be addressed to C.-R. C.

**Acknowledgement**

Thanks for NTU-IBM Q Hub at National Taiwan University from Ministry of Science and Technology, Taiwan, under grant No. MOST 107-2627-E-002-001-MY3, 108-2627-E-002-002 and NTU-107L104064 provide quantum computer resources.

**Supplementary Information**

**Supplement A**

For N-qubits, the recursive relation of Mermin's polynomials are

$$M_n = M_{n-1}(a_n + a'_n) + M'_{n-1}(a_n - a'_n) \quad (A.1)$$

where $M_1 = a_1 = X, M'_1 = a'_1 = Y, a_n = X, a'_n = Y$, and integer $n \geq 2$. Here $X$ and $Y$ represent Pauli-$X$ gate and Pauli-$Y$ gate respectively. For GHZ-like states with an adjustable phase angle $\varphi$ is

$$|\Psi_n\rangle = \frac{1}{\sqrt{2}}(|00\ldots0\rangle + e^{i\varphi}|11\ldots1\rangle) \quad (A.2)$$

When (A.1) operates on (A.2), we will obtain

$$M_n|\Psi_n\rangle = \frac{1}{\sqrt{2}}\left(\sum_{j=0}^{n} M_n^j X^{n-j} Y^j |00\ldots0\rangle + e^{i\varphi}\sum_{j=0}^{n} M_n^j X^{n-j} Y^j |11\ldots1\rangle\right)$$

$$= \frac{1}{\sqrt{2}}\left(e^{i\varphi}\sum_{j=0}^{n} M_n^j(-i)^j|00\ldots0\rangle + \sum_{j=0}^{n} M_n^j(i)^j|11\ldots1\rangle\right) \quad (A.3)$$

Then

$$\langle M_n\rangle = \langle\Psi_n|M_n|\Psi_n\rangle$$

$$= \frac{1}{2}\left[e^{i\varphi}\sum_{j=0}^{n} M_n^j(-i)^j + e^{-i\varphi}\sum_{j=0}^{n} M_n^j(i)^j\right]$$

$$= \frac{1}{2}\sum_{j=0}^{n} M_n^j \left[e^{i\left(\varphi - \frac{j\pi}{2}\right)} + e^{-i\left(\varphi - \frac{j\pi}{2}\right)}\right]$$

$$= \sum_{j=0}^{n} M_n^j \cos\left(\varphi - \frac{j\pi}{2}\right) \quad (A.4)$$

where $M_n^j = C_n^j f(n,j)$ and $C_n^j$ is the number of combinations, and $f(n,j)$ values are listed Table SA-I,

| $f(n, j\|s, r)$ | s=0 | s=1 | s=2 | s=3 | s=4 | s=5 | s=6 | s=7 |
|---|---|---|---|---|---|---|---|---|
| r=0 | 1 | 1 | 1 | 0 | -1 | -1 | -1 | 0 |
| r=1 | -1 | 0 | 1 | 1 | 1 | 0 | -1 | -1 |
| r=2 | -1 | -1 | -1 | 0 | 1 | 1 | 1 | 0 |
| r=3 | 1 | 0 | -1 | -1 | -1 | 0 | 1 | 1 |

Table SA-I: The table for $f(n,j|s,r)$ values. The remainders are $s = n \bmod 8$ and $r = j \bmod 4$.

For the N-qubits entangled system, $n \geq 2$. Putting $M_n^j = C_n^j f(n,j)$ values into (A.4), we have all expectations for $\langle M_n \rangle$:

If $s = n \bmod 8 = 2$,

$$\langle M_n \rangle = \sum_{j=0,4,8...}^{n-2} C_n^j \cos(\varphi) + \sum_{j=1,5,9...}^{n-1} C_n^j \cos\left(\varphi - \frac{\pi}{2}\right) - \sum_{j=2,6,10...}^{n} C_n^j \cos(\varphi - \pi)$$

$$- \sum_{j=3,7,11...}^{n-3} C_n^j \cos\left(\varphi - \frac{3\pi}{2}\right)$$

$$= \sum_{j=0,2,4...}^{n} C_n^j \cos(\varphi) + \sum_{j=1,3,5...}^{n-1} C_n^j \cos\left(\varphi - \frac{\pi}{2}\right)$$

$$= \sum_{j=0,1,2...}^{n-1} C_{n-1}^j \left[\cos(\varphi) + \cos\left(\varphi - \frac{\pi}{2}\right)\right]$$

$$= 2^{n-1} \times 2 \cos\left(\varphi - \frac{\pi}{4}\right) \cos\left(\frac{\pi}{4}\right)$$

$$= 2^{n-\frac{1}{2}} \cos\left(\varphi - \frac{\pi}{4}\right)$$

If $s = n \bmod 8 = 3$,

$$\langle M_n \rangle = \sum_{j=1,5,9...}^{n-2} C_n^j \cos\left(\varphi - \frac{\pi}{2}\right) - \sum_{j=3,7,11...}^{n} C_n^j \cos\left(\varphi - \frac{3\pi}{2}\right)$$

$$= \sum_{j=1,3,5...}^{n} C_n^j \cos\left(\varphi - \frac{\pi}{2}\right)$$

$$= \sum_{j=0,1,2...}^{n-1} C_{n-1}^j \cos\left(\varphi - \frac{\pi}{2}\right)$$

$$= 2^{n-1} \cos\left(\varphi - \frac{\pi}{2}\right)$$

If $s = n \bmod 8 = 4$,

$$\langle M_n \rangle = -\sum_{j=0,4,8...}^{n} C_n^j \cos(\varphi) + \sum_{j=1,5,9...}^{n-3} C_n^j \cos\left(\varphi - \frac{\pi}{2}\right) + \sum_{j=2,6,10...}^{n-2} C_n^j \cos(\varphi - \pi)$$

$$- \sum_{j=3,7,11...}^{n-1} C_n^j \cos\left(\varphi - \frac{3\pi}{2}\right)$$

$$= \sum_{j=0,2,4...}^{n} C_n^j \cos(\varphi - \pi) + \sum_{j=1,3,5...}^{n-1} C_n^j \cos\left(\varphi - \frac{\pi}{2}\right)$$

$$= \sum_{j=0,1,2...}^{n-1} C_{n-1}^j \left[\cos(\varphi - \pi) + \cos\left(\varphi - \frac{\pi}{2}\right)\right]$$

$$= 2^{n-1} \times 2 \cos\left(\varphi - \frac{3\pi}{4}\right) \cos\left(\frac{\pi}{4}\right)$$

$$= 2^{n-\frac{1}{2}} \cos\left(\varphi - \frac{3\pi}{4}\right)$$

If $s = n \bmod 8 = 5$,

$$\langle M_n \rangle = - \sum_{j=0,4,8...}^{n-1} C_n^j \cos(\varphi) + \sum_{j=2,6,10...}^{n-3} C_n^j \cos(\varphi - \pi)$$

$$= \sum_{j=0,2,4...}^{n-1} C_n^j \cos(\varphi - \pi)$$

$$= \sum_{j=0,1,2...}^{n-1} C_{n-1}^j \cos(\varphi - \pi)$$

$$= 2^{n-1} \cos(\varphi - \pi)$$

If $s = n \bmod 8 = 6$,

$$\langle M_n \rangle = - \sum_{j=0,4,8...}^{n-2} C_n^j \cos(\varphi) - \sum_{j=1,5,9...}^{n-1} C_n^j \cos\left(\varphi - \frac{\pi}{2}\right) + \sum_{j=2,6,10...}^{n} C_n^j \cos(\varphi - \pi)$$

$$+ \sum_{j=3,7,11...}^{n-3} C_n^j \cos\left(\varphi - \frac{3\pi}{2}\right)$$

$$= \sum_{j=0,2,4...}^{n} C_n^j \cos(\varphi - \pi) + \sum_{j=1,3,5...}^{n-1} C_n^j \cos\left(\varphi - \frac{3\pi}{2}\right)$$

$$= \sum_{j=0,1,2...}^{n-1} C_{n-1}^j \left[\cos(\varphi - \pi) + \cos\left(\varphi - \frac{3\pi}{2}\right)\right]$$

$$= 2^{n-1} \times 2 \cos\left(\varphi - \frac{5\pi}{4}\right) \cos\left(\frac{\pi}{4}\right)$$

$$= 2^{n-\frac{1}{2}} \cos\left(\varphi - \frac{5\pi}{4}\right)$$

If $s = n \bmod 8 = 7$,

$$\langle M_n \rangle = -\sum_{j=1,5,9\ldots}^{n-2} C_n^j \cos\left(\varphi - \frac{\pi}{2}\right) + \sum_{j=3,7,11\ldots}^{n} C_n^j \cos\left(\varphi - \frac{3\pi}{2}\right)$$

$$= \sum_{j=1,3,5\ldots}^{n} C_n^j \cos\left(\varphi - \frac{3\pi}{2}\right)$$

$$= \sum_{j=0,1,2\ldots}^{n-1} C_{n-1}^j \cos\left(\varphi - \frac{3\pi}{2}\right)$$

$$= 2^{n-1} \cos\left(\varphi - \frac{3\pi}{2}\right)$$

If $s = n \bmod 8 = 0$,

$$\langle M_n \rangle = \sum_{j=0,4,8\ldots}^{n} C_n^j \cos(\varphi) - \sum_{j=1,5,9\ldots}^{n-3} C_n^j \cos\left(\varphi - \frac{\pi}{2}\right) - \sum_{j=2,6,10\ldots}^{n-2} C_n^j \cos(\varphi - \pi)$$

$$+ \sum_{j=3,7,11\ldots}^{n-1} C_n^j \cos\left(\varphi - \frac{3\pi}{2}\right)$$

$$= -\sum_{j=0,2,4\ldots}^{n} C_n^j \cos\left(\varphi - \frac{\pi}{2}\right) - \sum_{j=1,3,5\ldots}^{n-1} C_n^j \cos(\varphi - \pi)$$

$$= -\sum_{j=0,1,2\ldots}^{n-1} C_{n-1}^j \left[\cos\left(\varphi - \frac{\pi}{2}\right) + \cos(\varphi - \pi)\right]$$

$$= 2^{n-1} \times 2\cos\left(\varphi - \frac{7\pi}{4}\right)\cos\left(\frac{\pi}{4}\right)$$

$$= 2^{n-\frac{1}{2}} \cos\left(\varphi - \frac{7\pi}{4}\right)$$

If $s = n \bmod 8 = 1$,

$$\langle M_n \rangle = \sum_{j=0,4,8\ldots}^{n-1} C_n^j \cos(\varphi) - \sum_{j=2,6,10\ldots}^{n-3} C_n^j \cos(\varphi - \pi)$$

$$= \sum_{j=0,2,4\ldots}^{n-1} C_n^j \cos(\varphi)$$

$$= \sum_{j=0,1,2\ldots}^{n-1} C_{n-1}^j \cos(\varphi)$$

$$= 2^{n-1} \cos(\varphi)$$

In summary, the above results of $\langle M_n \rangle$ can be deduced into

$$\langle M_n \rangle = \begin{cases} 2^{n-\frac{1}{2}} \cos\left(\varphi - \frac{n-1}{4}\pi\right), n = even \\ 2^{n-1} \cos\left(\varphi - \frac{n-1}{4}\pi\right), n = odd \end{cases}$$

Similarly, it can be proven that

$$\langle M_n' \rangle = \begin{cases} 2^{n-\frac{1}{2}} \sin\left(\varphi - \frac{n-1}{4}\pi\right), n = even \\ 2^{n-1} \sin\left(\varphi - \frac{n-1}{4}\pi\right), n = odd \end{cases}$$

**Supplement B**

Table for sequence number of the connectivity of N-qubits and the orthogonal measurements of $\langle M_n \rangle$ and $\langle M'_n \rangle$.

(SB-I) 2 qubits

| No. | qubits | $\langle M \rangle$ $=\sum \frac{M}{5}$ | $\langle M' \rangle$ $=\sum \frac{M'}{5}$ | $\sigma_M$ $=\sqrt{\frac{\sum(M-\langle M \rangle)^2}{5}}$ | $\sigma_{M'}$ $=\sqrt{\frac{\sum(M'-\langle M' \rangle)^2}{5}}$ |
|---|---|---|---|---|---|
| 1 | [29, 36] | 2.3129 | -0.0285 | 0.2526 | 0.0911 |
| 2 | [33, 34] | 2.284 | 0.0629 | 0.1313 | 0.1249 |
| 3 | [38, 41] | 2.2809 | 0.0926 | 0.1078 | 0.0947 |
| 4 | [7, 16] | 2.2766 | 0.0344 | 0.0874 | 0.0322 |
| 5 | [25, 26] | 2.243 | -0.2211 | 0.0571 | 0.0947 |
| 6 | [19, 20] | 2.2184 | -0.2637 | 0.0484 | 0.088 |
| 7 | [40, 46] | 2.1285 | 0.1668 | 0.0651 | 0.1532 |
| 8 | [11, 17] | 2.1141 | -0.0391 | 0.1166 | 0.1163 |
| 9 | [16, 19] | 2.1133 | -0.207 | 0.061 | 0.1921 |
| 10 | [20, 21] | 2.1125 | 0.1422 | 0.0301 | 0.2126 |
| 11 | [18, 27] | 2.1078 | -0.0563 | 0.1722 | 0.2487 |
| 12 | [34, 40] | 2.102 | 0.0184 | 0.0721 | 0.0957 |
| 13 | [36, 37] | 2.0586 | 0.1273 | 0.0546 | 0.0645 |
| 14 | [0, 1] | 2.018 | -0.2945 | 0.4581 | 0.4831 |
| 15 | [3, 4] | 1.9684 | -0.0309 | 0.2576 | 0.3528 |
| 16 | [4, 6] | 1.9598 | 0.2762 | 0.0988 | 0.2004 |
| 17 | [26, 27] | 1.9293 | -0.0418 | 0.1903 | 0.1706 |
| 18 | [46, 47] | 1.925 | -0.025 | 0.0642 | 0.0608 |
| 19 | [41, 50] | 1.9105 | -0.4402 | 0.1677 | 0.279 |
| 20 | [48, 49] | 1.8375 | 0.1125 | 0.4107 | 0.0849 |
| 21 | [45, 46] | 1.8137 | 0.0348 | 0.0509 | 0.1855 |
| 22 | [49, 50] | 1.7719 | -0.25 | 0.0977 | 0.2302 |
| 23 | [21, 22] | 1.7672 | -0.1219 | 0.108 | 0.1258 |
| 24 | [30, 31] | 1.7594 | -0.0828 | 0.1448 | 0.5188 |
| 25 | [44, 51] | 1.7242 | 0.1789 | 0.1263 | 0.1392 |
| 26 | [47, 48] | 1.7047 | 0.0266 | 0.3335 | 0.0432 |
| 27 | [11, 12] | 1.6648 | -0.0102 | 0.1746 | 0.3402 |
| 28 | [21, 28] | 1.6395 | -0.3301 | 0.079 | 0.1602 |
| 29 | [39, 42] | 1.607 | -0.0148 | 0.1849 | 0.1523 |
| 30 | [32, 33] | 1.5906 | 0.1203 | 0.1717 | 0.1208 |
| 31 | [14, 15] | 1.5742 | 0.043 | 0.0815 | 0.0541 |
| 32 | [22, 23] | 1.5699 | -0.1902 | 0.1673 | 0.1964 |
| 33 | [34, 35] | 1.552 | 0.1762 | 0.0897 | 0.1482 |
| 34 | [35, 36] | 1.5359 | 0.0578 | 0.0556 | 0.2299 |
| 35 | [10, 11] | 1.5297 | -0.1906 | 0.1001 | 0.1488 |
| 36 | [31, 32] | 1.4918 | 0.1785 | 0.1144 | 0.1075 |
| 37 | [9, 10] | 1.4801 | -0.1004 | 0.07 | 0.0791 |
| 38 | [24, 25] | 1.4684 | 1.0191 | 0.4552 | 0.5552 |
| 39 | [17, 23] | 1.4332 | -0.4988 | 0.1397 | 0.229 |
| 40 | [25, 29] | 1.4059 | 1.2348 | 0.0904 | 0.0816 |
| 41 | [44, 45] | 1.3781 | 0.0813 | 0.1399 | 0.0962 |
| 42 | [0, 5] | 1.3625 | -0.0203 | 0.1706 | 0.0839 |
| 43 | [30, 39] | 1.3422 | 0.2094 | 0.1651 | 0.2213 |
| 44 | [37, 38] | 1.2305 | -0.3086 | 0.1779 | 0.1373 |
| 45 | [7, 8] | 1.1336 | -0.068 | 0.1117 | 0.1645 |
| 46 | [15, 18] | 1.1145 | 1.123 | 0.0421 | 0.0471 |
| 47 | [28, 32] | 1.1027 | 0.2395 | 0.3294 | 0.3412 |
| 48 | [5, 9] | 1.0996 | 0.0223 | 0.1676 | 0.0995 |
| 49 | [8, 9] | 1.0082 | -0.0395 | 0.1023 | 0.0554 |
| 50 | [42, 43] | 0.9781 | 0.0688 | 0.0897 | 0.0702 |
| 51 | [48, 52] | 0.9555 | 0.0461 | 0.3487 | 0.3053 |
| 52 | [43, 44] | 0.8781 | 0.0063 | 0.0452 | 0.0575 |
| 53 | [23, 24] | 0.7285 | 0.1277 | 0.7465 | 0.0976 |
| 54 | [6, 13] | 0.6656 | 0.025 | 0.0853 | 0.0732 |
| 55 | [1, 2] | 0.6359 | -0.0922 | 0.0886 | 0.1334 |
| 56 | [2, 3] | 0.5887 | -0.0309 | 0.1171 | 0.1716 |
| 57 | [12, 13] | 0.5191 | -0.002 | 0.0618 | 0.1112 |
| 58 | [13, 14] | 0.5148 | 0.0445 | 0.095 | 0.1178 |

(SB-II) 3 qubits

| No. | qubits | $\langle M \rangle = \sum \frac{M}{5}$ | $\langle M' \rangle = \sum \frac{M'}{5}$ | $\sigma_M = \sqrt{\frac{\sum(M-\langle M \rangle)^2}{5}}$ | $\sigma_{M'} = \sqrt{\frac{\sum(M'-\langle M' \rangle)^2}{5}}$ |
|---|---|---|---|---|---|
| 1 | [16, 19, 20] | 2.8352 | 0.0895 | 0.0501 | 0.3988 |
| 2 | [7, 16, 19] | 2.7836 | 0.1742 | 0.0971 | 0.0861 |
| 3 | [24, 25, 26] | 2.7012 | 0.2016 | 0.0328 | 0.1138 |
| 4 | [24, 25, 29] | 2.6969 | 0.0426 | 0.0587 | 0.1334 |
| 5 | [46, 47, 48] | 2.6082 | -0.1422 | 0.1213 | 0.2602 |
| 6 | [33, 34, 40] | 2.4199 | 0.5195 | 0.3514 | 0.1997 |
| 7 | [19, 20, 21] | 2.3922 | -0.0285 | 0.1029 | 0.2533 |
| 8 | [25, 26, 27] | 2.3082 | 0.1234 | 0.2536 | 0.4073 |
| 9 | [47, 48, 49] | 2.2844 | -0.2602 | 0.0933 | 0.149 |
| 10 | [26, 25, 29] | 2.2641 | 0.0105 | 0.0626 | 0.2109 |
| 11 | [38, 41, 50] | 2.2609 | 0.3375 | 0.2196 | 0.3567 |
| 12 | [18, 27, 26] | 2.1449 | 0.541 | 0.1563 | 0.3539 |
| 13 | [41, 50, 49] | 2.1418 | 0.0539 | 0.1717 | 0.1505 |
| 14 | [48, 49, 50] | 2.073 | 0.4699 | 0.3031 | 0.1905 |
| 15 | [8, 7, 16] | 2.0723 | 0.875 | 0.8219 | 1.0645 |
| 16 | [12, 11, 17] | 2.0504 | -0.2855 | 0.2755 | 0.3127 |
| 17 | [20, 21, 28] | 1.9918 | 0.3387 | 0.0934 | 0.1425 |
| 18 | [45, 46, 47] | 1.9902 | 0.482 | 0.16 | 0.2163 |
| 19 | [20, 21, 22] | 1.9691 | -0.3785 | 0.1109 | 0.2012 |
| 20 | [17, 23, 24] | 1.9641 | 0.5152 | 0.1763 | 0.5863 |
| 21 | [22, 23, 24] | 1.927 | 0.4766 | 0.1289 | 0.5209 |
| 22 | [25, 29, 36] | 1.9227 | 0.6891 | 0.1343 | 0.1158 |
| 23 | [15, 18, 27] | 1.918 | 0.5488 | 0.1956 | 0.1242 |
| 24 | [40, 46, 47] | 1.9136 | 0.52 | 0.1057 | 0.2497 |
| 25 | [33, 34, 35] | 1.8965 | 0.3059 | 0.1412 | 0.1141 |
| 26 | [3, 4, 6] | 1.877 | 0.2031 | 0.7382 | 0.2926 |
| 27 | [17, 23, 22] | 1.7773 | 0.616 | 0.2471 | 0.4719 |
| 28 | [7, 8, 9] | 1.7566 | 0.375 | 0.7507 | 0.8332 |
| 29 | [40, 46, 45] | 1.7476 | 0.1797 | 0.07 | 0.086 |
| 30 | [34, 35, 36] | 1.6914 | 0.175 | 0.1344 | 0.3035 |
| 31 | [22, 21, 28] | 1.6684 | 0.2375 | 0.1773 | 0.2955 |
| 32 | [32, 33, 34] | 1.6453 | 0.4055 | 0.2808 | 0.2393 |
| 33 | [23, 24, 25] | 1.6367 | 0.3176 | 0.4162 | 1.5225 |
| 34 | [31, 30, 39] | 1.6117 | 0.0348 | 0.1258 | 0.3336 |
| 35 | [44, 45, 46] | 1.5934 | -0.1664 | 0.2217 | 0.853 |
| 36 | [8, 9, 10] | 1.5836 | 0.4262 | 0.1529 | 0.5078 |
| 37 | [31, 32, 33] | 1.5758 | 0.0117 | 0.6335 | 0.1724 |
| 38 | [29, 36, 35] | 1.5637 | -0.0867 | 0.0414 | 0.147 |
| 39 | [10, 11, 17] | 1.5625 | -0.0668 | 0.1801 | 0.1235 |
| 40 | [35, 34, 40] | 1.5344 | -0.0246 | 0.157 | 0.6078 |
| 41 | [49, 48, 52] | 1.5254 | 0.2764 | 0.2951 | 0.2313 |
| 42 | [45, 44, 51] | 1.4795 | 0.0586 | 0.3042 | 0.2753 |
| 43 | [14, 15, 18] | 1.4703 | 0.3941 | 0.1175 | 0.1435 |
| 44 | [9, 10, 11] | 1.4375 | 0.352 | 0.1563 | 0.1233 |
| 45 | [30, 31, 32] | 1.4359 | 0.2633 | 0.0668 | 0.1882 |
| 46 | [30, 39, 42] | 1.4242 | 0.0711 | 0.0597 | 0.2339 |
| 47 | [21, 22, 23] | 1.4027 | 0.5035 | 0.0786 | 0.1102 |
| 48 | [47, 48, 52] | 1.3931 | -0.5444 | 0.1013 | 0.2702 |
| 49 | [29, 36, 37] | 1.3844 | 0.4313 | 0.1998 | 0.1836 |
| 50 | [37, 38, 41] | 1.3809 | 0.1953 | 0.1003 | 0.1644 |
| 51 | [10, 11, 12] | 1.3773 | 0.1961 | 0.1143 | 0.2885 |
| 52 | [28, 32, 31] | 1.3559 | 0.2875 | 0.1353 | 0.1877 |
| 53 | [36, 37, 38] | 1.3527 | 0.3242 | 0.1199 | 0.1145 |
| 54 | [42, 43, 44] | 1.2574 | -0.1316 | 0.0815 | 0.1759 |
| 55 | [0, 5, 9] | 1.2336 | 0.0914 | 0.1962 | 0.1586 |
| 56 | [28, 32, 33] | 1.2262 | 0.2488 | 0.326 | 0.2061 |
| 57 | [21, 28, 32] | 1.2074 | -0.1445 | 0.1064 | 0.202 |
| 58 | [35, 36, 37] | 1.2035 | 0.548 | 0.0772 | 0.1882 |
| 59 | [11, 17, 23] | 1.1918 | 0.6965 | 0.8115 | 1.0617 |
| 60 | [39, 42, 43] | 1.1137 | 0.0773 | 0.1122 | 0.097 |
| 61 | [1, 0, 5] | 1.0957 | 0.1801 | 0.3018 | 0.152 |
| 62 | [43, 44, 51] | 1.0023 | -0.0496 | 0.1177 | 0.2259 |
| 63 | [43, 44, 45] | 0.9633 | -0.0418 | 0.0816 | 0.1758 |
| 64 | [5, 9, 8] | 0.8004 | 0.2637 | 0.1737 | 0.2076 |

| No. | qubits | | | | |
|---|---|---|---|---|---|
| 65 | [4, 6, 13] | 0.7906 | 0.1949 | 0.1592 | 0.0987 |
| 66 | [5, 9, 10] | 0.7727 | 0.1422 | 0.1355 | 0.1186 |
| 67 | [2, 3, 4] | 0.7176 | 0.0387 | 0.0463 | 0.2174 |
| 68 | [11, 12, 13] | 0.7125 | 0.048 | 0.2051 | 0.1626 |
| 69 | [6, 13, 12] | 0.6938 | 0.1695 | 0.0952 | 0.3154 |
| 70 | [0, 1, 2] | 0.6914 | 0.0609 | 0.1881 | 0.0461 |
| 71 | [1, 2, 3] | 0.6395 | -0.0047 | 0.0498 | 0.2018 |
| 72 | [6, 13, 14] | 0.5016 | 0.1949 | 0.1447 | 0.194 |
| 73 | [13, 14, 15] | 0.4938 | 0.0844 | 0.0667 | 0.1631 |
| 74 | [12, 13, 14] | 0.416 | 0.0215 | 0.1082 | 0.1049 |
| 75 | [34, 40, 41] | 0.3926 | 0.0254 | 0.4105 | 0.262 |

(SB-III) 4 qubits

| No. | qubits | $\langle M \rangle = \sum \frac{M}{5}$ | $\langle M' \rangle = \sum \frac{M'}{5}$ | $\sigma_M = \sqrt{\frac{\sum(M-\langle M \rangle)^2}{5}}$ | $\sigma_{M'} = \sqrt{\frac{\sum(M'-\langle M' \rangle)^2}{5}}$ |
|---|---|---|---|---|---|
| 1 | [7, 16, 19, 20] | 3.6434 | 0.9137 | 0.0698 | 0.1626 |
| 2 | [16, 19, 20, 21] | 3.0605 | 0.9902 | 0.1926 | 0.3331 |
| 3 | [18, 27, 26, 25] | 3.0217 | 0.9717 | 0.2938 | 0.4996 |
| 4 | [17, 23, 24, 25] | 2.6338 | 0.2885 | 0.1602 | 0.1185 |
| 5 | [41, 50, 49, 48] | 2.6217 | 0.5764 | 0.2436 | 0.6053 |
| 6 | [22, 23, 24, 25] | 2.4162 | 1.0068 | 0.1016 | 0.4257 |
| 7 | [23, 24, 25, 26] | 2.4104 | 0.4713 | 0.2365 | 0.5264 |
| 8 | [38, 41, 50, 49] | 2.3688 | 0.6656 | 0.4105 | 0.3042 |
| 9 | [19, 20, 21, 22] | 2.3451 | -0.1689 | 0.1125 | 0.1847 |
| 10 | [17, 11, 10, 9] | 2.3104 | 0.4932 | 0.0522 | 0.2527 |
| 11 | [21, 20, 22, 28] | 2.2922 | 0.1141 | 0.1194 | 0.1741 |
| 12 | [46, 47, 48, 49] | 2.2904 | 0.2076 | 0.213 | 0.5762 |
| 13 | [28, 21, 20, 19] | 2.2217 | 0.7154 | 0.1374 | 0.2017 |
| 14 | [11, 17, 23, 22] | 2.2154 | -0.0205 | 0.1625 | 0.1559 |
| 15 | [17, 23, 22, 21] | 2.2078 | 0.4266 | 0.1976 | 0.2365 |
| 16 | [40, 46, 47, 48] | 2.1738 | 0.6895 | 0.3539 | 0.4128 |
| 17 | [47, 48, 49, 50] | 2.0277 | 0.334 | 0.1093 | 0.2423 |
| 18 | [15, 18, 27, 26] | 2.0244 | 1.1275 | 0.1732 | 0.3441 |
| 19 | [23, 17, 22, 24] | 1.8908 | 0.0393 | 0.1562 | 0.2991 |
| 20 | [10, 11, 17, 23] | 1.8355 | 0.4512 | 0.1169 | 0.0497 |
| 21 | [34, 33, 35, 40] | 1.8137 | 0.7715 | 0.1614 | 0.3227 |
| 22 | [11, 10, 12, 17] | 1.8012 | 0.5418 | 0.3005 | 0.1429 |
| 23 | [45, 46, 47, 48] | 1.7551 | 0.3207 | 0.444 | 0.3864 |
| 24 | [21, 22, 23, 24] | 1.7252 | 1.6674 | 0.1639 | 0.2652 |
| 25 | [33, 34, 40, 46] | 1.7139 | 1.7795 | 0.1721 | 0.3667 |
| 26 | [12, 11, 17, 23] | 1.6785 | 0.1848 | 0.3513 | 0.413 |
| 27 | [37, 38, 41, 50] | 1.6742 | 0.9445 | 0.367 | 0.1655 |
| 28 | [40, 34, 33, 32] | 1.6531 | 0.7359 | 0.2131 | 0.2315 |
| 29 | [52, 48, 47, 46] | 1.5852 | 0.4867 | 0.3282 | 0.3904 |

| No. | qubits | ⟨M⟩ | ⟨M'⟩ | σ_M | σ_M' |
|---|---|---|---|---|---|
| 30 | [52, 48, 49, 50] | 1.5383 | 0.4336 | 0.095 | 0.2118 |
| 31 | [20, 21, 22, 23] | 1.5205 | 1.1549 | 0.0832 | 0.0766 |
| 32 | [20, 21, 28, 32] | 1.4814 | 1.0424 | 0.0375 | 0.2278 |
| 33 | [8, 7, 16, 19] | 1.4631 | 0.4412 | 0.0782 | 0.2281 |
| 34 | [48, 47, 49, 52] | 1.1508 | 0.568 | 0.3334 | 0.3927 |
| 35 | [17, 11, 12, 13] | 0.8455 | 0.3893 | 0.153 | 0.2241 |
| 36 | [6, 4, 3, 2] | 0.6705 | 0.1549 | 0.1472 | 0.0933 |
| 37 | [3, 4, 6, 13] | 0.5779 | 0.3967 | 0.246 | 0.1717 |
| 38 | [29, 25, 24, 23] | 0.1854 | 0.4838 | 0.453 | 0.1406 |

(SB-IV) 5 qubits

| No. | qubits | $\langle M \rangle = \sum \frac{M}{5}$ | $\langle M' \rangle = \sum \frac{M'}{5}$ | $\sigma_M = \sqrt{\frac{\sum (M - \langle M \rangle)^2}{5}}$ | $\sigma_{M'} = \sqrt{\frac{\sum (M' - \langle M' \rangle)^2}{5}}$ |
|---|---|---|---|---|---|
| 1 | [18, 27, 26, 25, 29] | 2.9463 | 1.3346 | 0.3172 | 0.4767 |
| 2 | [17, 23, 24, 25, 26] | 2.8955 | -0.234 | 1.5937 | 1.8391 |
| 3 | [38, 41, 50, 49, 48] | 2.8201 | 1.2143 | 0.5122 | 0.6624 |
| 4 | [8, 7, 16, 19, 20] | 2.7723 | 1.1574 | 1.2694 | 1.4705 |
| 5 | [7, 16, 19, 20, 21] | 2.7172 | 1.7732 | 0.483 | 0.3485 |
| 6 | [17, 23, 24, 25, 29] | 2.6445 | -0.4066 | 1.0674 | 2.1041 |
| 7 | [41, 50, 49, 48, 47] | 2.6248 | 0.393 | 0.2861 | 1.2676 |
| 8 | [12, 11, 17, 23, 24] | 2.6094 | 0.032 | 0.6586 | 1.3528 |
| 9 | [18, 27, 26, 25, 24] | 2.532 | 1.7607 | 0.5603 | 0.7339 |

| 10 | [17, 23, 22, 21, 20] | 2.3252 | -0.1342 | 0.8296 | 1.5851 |
| 11 | [16, 19, 20, 21, 22] | 2.1982 | 0.7438 | 0.7345 | 0.6945 |
| 12 | [45, 46, 47, 48, 49] | 2.1809 | 1.1436 | 0.5013 | 0.6626 |
| 13 | [34, 40, 46, 47, 48] | 2.1596 | 0.9865 | 0.091 | 0.4624 |
| 14 | [46, 40, 45, 47, 48] | 2.1217 | 0.6338 | 0.2074 | 0.4041 |
| 15 | [22, 23, 24, 25, 26] | 2.1074 | 0.8063 | 0.1279 | 1.4282 |
| 16 | [46, 47, 48, 49, 50] | 2.0898 | 0.9191 | 0.5195 | 1.1295 |
| 17 | [16, 19, 20, 21, 28] | 1.9529 | 1.0408 | 0.5962 | 0.9444 |
| 18 | [21, 28, 22, 20, 19] | 1.907 | 1.3711 | 0.4331 | 0.1849 |
| 19 | [40, 46, 47, 48, 49] | 1.8965 | 1.6295 | 0.4579 | 0.399 |
| 20 | [12, 11, 17, 23, 22] | 1.8477 | -0.1818 | 0.5282 | 0.9377 |
| 21 | [22, 23, 24, 25, 29] | 1.8398 | 1.1664 | 0.3675 | 1.8413 |
| 22 | [44, 45, 46, 47, 48] | 1.6973 | 1.2568 | 0.4776 | 0.5376 |
| 23 | [23, 17, 22, 24, 25] | 1.6498 | 0.1639 | 0.4884 | 0.3427 |
| 24 | [45, 46, 47, 48, 52] | 1.6027 | 0.6314 | 0.3886 | 0.5596 |
| 25 | [10, 11, 17, 23, 22] | 1.5992 | 0.4705 | 0.3287 | 0.9452 |
| 26 | [17, 23, 22, 21, 28] | 1.5471 | -0.2238 | 0.8506 | 1.456 |
| 27 | [48, 49, 52, 47, 46] | 1.4859 | 0.7301 | 0.3662 | 0.4667 |
| 28 | [19, 20, 21, 28, 32] | 1.4391 | 0.7012 | 0.2019 | 0.5187 |
| 29 | [48, 52, 47, 49, 50] | 1.3164 | 0.443 | 0.4117 | 0.9234 |
| 30 | [11, 17, 23, 22, 21] | 1.3047 | -0.0078 | 0.5205 | 1.5196 |
| 31 | [11, 17, 12, 10, 9] | 1.3045 | 0.1677 | 0.2715 | 0.4286 |
| 32 | [11, 17, 23, 24, 25] | 1.2818 | -0.6475 | 1.1029 | 2.2129 |
| 33 | [41, 50, 49, 48, 52] | 1.2775 | 0.3967 | 0.4205 | 0.9731 |
| 34 | [8, 9, 10, 11, 17] | 1.2305 | 0.8055 | 0.7973 | 1.2579 |
| 35 | [23, 22, 24, 17, 11] | 1.2164 | 0.6332 | 0.1563 | 0.3045 |
| 36 | [21, 20, 28, 22, 23] | 1.1445 | 1.0697 | 0.0343 | 0.4258 |
| 37 | [23, 24, 25, 26, 27] | 1.0744 | 0.8533 | 0.5238 | 2.1887 |
| 38 | [23, 24, 17, 22, 21] | 1.0738 | 0.3967 | 0.1844 | 0.3654 |
| 39 | [11, 12, 10, 17, 23] | 1.0678 | 0.3533 | 0.2623 | 0.4457 |
| 40 | [15, 18, 27, 26, 25] | 1.0439 | 1.1283 | 0.3463 | 0.9969 |
| 41 | [19, 20, 21, 22, 23] | 1.0375 | 1.2807 | 0.2691 | 0.6667 |
| 42 | [40, 46, 47, 48, 52] | 0.9646 | 0.9105 | 0.701 | 0.7589 |
| 43 | [25, 29, 24, 26, 27] | 0.8465 | 0.6158 | 0.2664 | 0.5486 |
| 44 | [14, 15, 18, 27, 26] | 0.7588 | 0.7727 | 0.4976 | 0.5879 |
| 45 | [21, 22, 20, 28, 32] | 0.6242 | 0.9898 | 0.2624 | 0.3034 |
| 46 | [5, 9, 10, 11, 17] | 0.6082 | 0.7994 | 0.3819 | 0.2901 |
| 47 | [36, 37, 38, 41, 50] | 0.5998 | 1.1932 | 0.5714 | 0.5 |
| 48 | [11, 10, 17, 12, 13] | 0.5861 | 0.1275 | 0.0953 | 0.1361 |
| 49 | [9, 10, 11, 17, 23] | 0.5102 | 0.5061 | 0.3833 | 0.453 |
| 50 | [21, 22, 23, 24, 25] | 0.5098 | 1.3777 | 0.7494 | 0.5943 |
| 51 | [13, 12, 11, 17, 23] | 0.4811 | 0.1131 | 0.2879 | 0.3105 |
| 52 | [37, 38, 41, 50, 49] | 0.385 | 1.048 | 1.0614 | 0.2558 |

(SB-V) 6 qubits

| No. | qubits | $\langle M \rangle = \sum \frac{M}{5}$ | $\langle M' \rangle = \sum \frac{M'}{5}$ | $\sigma_M = \sqrt{\frac{\sum(M-\langle M \rangle)^2}{5}}$ | $\sigma_{M'} = \sqrt{\frac{\sum(M'-\langle M' \rangle)^2}{5}}$ |
|---|---|---|---|---|---|
| 1 | [23, 22, 17, 11, 24, 25] | 2.6616 | 0.838 | 0.85 | 0.5012 |
| 2 | [11, 17, 23, 22, 21, 20] | 2.3923 | -0.3165 | 0.4672 | 0.3914 |
| 3 | [17, 23, 22, 21, 20, 19] | 2.3797 | 1.2797 | 0.3603 | 0.7485 |
| 4 | [17, 23, 24, 25, 26, 27] | 2.3595 | 0.8085 | 0.4192 | 0.0894 |
| 5 | [23, 17, 22, 21, 24, 25] | 2.3353 | 0.6003 | 0.1862 | 0.3259 |
| 6 | [11, 17, 23, 24, 25, 26] | 2.1476 | -0.2515 | 0.1637 | 0.9041 |
| 7 | [23, 17, 22, 24, 25, 26] | 2.0757 | 0.6767 | 0.3653 | 0.8497 |
| 8 | [11, 17, 23, 24, 25, 29] | 1.8989 | 0.0304 | 0.5766 | 1.088 |
| 9 | [17, 23, 24, 25, 29, 36] | 1.8944 | 1.0185 | 0.4989 | 0.7976 |
| 10 | [7, 16, 19, 20, 21, 22] | 1.8688 | 1.0257 | 0.4488 | 1.9444 |
| 11 | [8, 7, 16, 19, 20, 21] | 1.7369 | 1.2092 | 0.7955 | 1.1109 |
| 12 | [23, 17, 22, 24, 25, 29] | 1.6824 | 0.959 | 0.6269 | 0.7215 |
| 13 | [11, 12, 10, 9, 17, 23] | 1.5885 | 0.8936 | 0.5259 | 0.1805 |
| 14 | [12, 11, 17, 23, 24, 25] | 1.5178 | 0.4287 | 0.4489 | 0.6955 |
| 15 | [7, 16, 19, 20, 21, 28] | 1.4373 | 1.2592 | 0.8469 | 1.5946 |
| 16 | [23, 24, 17, 22, 21, 20] | 1.3233 | 1.2188 | 0.487 | 0.6961 |
| 17 | [21, 28, 22, 20, 19, 16] | 1.249 | 1.7725 | 0.9191 | 0.575 |
| 18 | [11, 17, 23, 22, 21, 28] | 1.2233 | -0.4483 | 0.1742 | 0.4838 |
| 19 | [23, 22, 24, 17, 11, 12] | 1.2224 | 0.8706 | 0.7538 | 0.4393 |
| 20 | [11, 12, 10, 17, 23, 22] | 1.193 | 0.8121 | 0.3846 | 0.4537 |
| 21 | [11, 17, 12, 10, 9, 8] | 1.164 | 0.4028 | 0.2521 | 0.2219 |
| 22 | [10, 11, 17, 23, 24, 25] | 1.138 | 0.4687 | 0.5442 | 1.0308 |
| 23 | [11, 12, 10, 17, 23, 24] | 1.1326 | 1.4182 | 0.1786 | 0.4794 |
| 24 | [22, 23, 24, 25, 26, 27] | 1.1182 | 1.8283 | 1.3783 | 0.6776 |
| 25 | [41, 50, 49, 48, 47, 46] | 1.0979 | 0.7313 | 0.8666 | 0.5771 |
| 26 | [46, 40, 45, 47, 48, 49] | 1.0696 | 1.1597 | 0.6261 | 0.3333 |
| 27 | [10, 11, 17, 23, 22, 21] | 1.0627 | 0.5412 | 0.1342 | 0.4005 |
| 28 | [38, 41, 50, 49, 48, 47] | 1.0354 | 1.4494 | 0.6379 | 0.4395 |
| 29 | [12, 11, 17, 23, 22, 21] | 1.0347 | -0.1851 | 0.4502 | 0.74 |
| 30 | [23, 22, 24, 17, 11, 10] | 0.9272 | 0.6767 | 0.6491 | 0.2836 |
| 31 | [18, 27, 26, 25, 29, 36] | 0.9118 | 1.7729 | 0.2894 | 0.2727 |
| 32 | [11, 17, 12, 10, 9, 5] | 0.9061 | 0.5186 | 0.1457 | 0.3111 |
| 33 | [35, 34, 40, 46, 47, 48] | 0.8162 | 1.1799 | 0.6469 | 0.4416 |
| 34 | [38, 41, 50, 49, 48, 52] | 0.8004 | 1.3816 | 0.7294 | 0.6266 |
| 35 | [11, 17, 12, 13, 10, 9] | 0.7743 | 0.1979 | 0.1582 | 0.3601 |
| 36 | [15, 18, 27, 26, 25, 29] | 0.7176 | 1.4531 | 0.4886 | 0.4597 |
| 37 | [48, 47, 52, 49, 50, 41] | 0.7159 | 1.4177 | 0.0519 | 0.1922 |
| 38 | [9, 10, 11, 17, 23, 22] | 0.7086 | 1.0758 | 0.2935 | 0.1852 |
| 39 | [16, 19, 20, 21, 28, 32] | 0.6223 | 0.8066 | 0.7928 | 1.2625 |
| 40 | [25, 24, 29, 26, 27, 18] | 0.5567 | 0.6956 | 0.4145 | 0.7898 |
| 41 | [40, 46, 47, 48, 49, 50] | 0.4248 | 1.9201 | 0.6828 | 0.2015 |

| No. | qubits | ⟨M⟩ | ⟨M'⟩ | σ_M | σ_M' |
|---|---|---|---|---|---|
| 42 | [14, 15, 18, 27, 26, 25] | 0.408 | 1.2205 | 0.4045 | 0.3125 |
| 43 | [37, 38, 41, 50, 49, 48] | 0.2807 | 1.8057 | 0.5359 | 0.246 |
| 44 | [25, 29, 26, 24, 23, 22] | 0.2587 | 0.4413 | 0.2371 | 0.5079 |
| 45 | [34, 40, 46, 47, 48, 45] | 0.2437 | 0.1438 | 0.2944 | 0.5137 |
| 46 | [48, 52, 49, 47, 46, 40] | 0.2422 | 1.0234 | 0.4732 | 0.6656 |
| 47 | [13, 12, 11, 17, 23, 24] | 0.184 | 0.3703 | 0.3302 | 0.3959 |
| 48 | [15, 18, 27, 26, 25, 24] | 0.1368 | 1.6237 | 0.3413 | 0.4172 |
| 49 | [13, 12, 11, 17, 23, 22] | 0.1192 | 0.0804 | 0.3977 | 0.0905 |
| 50 | [18, 27, 26, 25, 24, 23] | 0.0831 | 1.4024 | 0.2246 | 0.4648 |
| 51 | [25, 29, 26, 24, 23, 17] | 0.0597 | 0.5497 | 0.2713 | 0.1521 |
| 52 | [34, 40, 46, 47, 48, 49] | 0.0494 | 1.3754 | 0.3026 | 0.5648 |
| 53 | [16, 19, 20, 21, 22, 23] | -0.1811 | 1.1346 | 1.152 | 1.5575 |
| 54 | [33, 34, 40, 46, 47, 48] | -0.4501 | 2.1396 | 0.1837 | 0.4812 |
| 55 | [21, 22, 23, 24, 25, 26] | -0.5388 | 1.6501 | 0.5097 | 0.3172 |

(SB-VI) 7 qubits

| No. | qubits | $\langle M \rangle = \sum \frac{M}{5}$ | $\langle M' \rangle = \sum \frac{M'}{5}$ | $\sigma_M = \sqrt{\frac{\sum (M - \langle M \rangle)^2}{5}}$ | $\sigma_{M'} = \sqrt{\frac{\sum (M' - \langle M' \rangle)^2}{5}}$ |
|---|---|---|---|---|---|
| 1 | [23, 17, 11, 22, 21, 24, 25] | 2.9992 | 1.3113 | 0.3936 | 0.535 |
| 2 | [23, 17, 22, 21, 24, 25, 26] | 2.4354 | 0.5891 | 0.5656 | 0.6591 |
| 3 | [23, 17, 22, 21, 24, 25, 29] | 2.3884 | 0.6784 | 0.3431 | 0.2396 |
| 4 | [23, 17, 24, 25, 22, 21, 20] | 2.3847 | 1.0421 | 0.4363 | 0.6337 |
| 5 | [23, 22, 17, 11, 24, 25, 26] | 1.9128 | 1.4679 | 0.6996 | 0.54 |
| 6 | [23, 17, 24, 25, 22, 21, 28] | 1.8985 | 0.2987 | 0.4706 | 0.4469 |
| 7 | [23, 22, 24, 25, 17, 11, 12] | 1.8346 | 1.247 | 0.6473 | 0.4866 |
| 8 | [12, 11, 17, 23, 24, 25, 26] | 1.8337 | 0.3945 | 0.7155 | 0.8928 |
| 9 | [23, 24, 17, 11, 22, 21, 20] | 1.6281 | 1.314 | 0.4539 | 0.6954 |
| 10 | [11, 17, 23, 24, 25, 29, 36] | 1.5223 | 0.1227 | 0.6953 | 0.702 |
| 11 | [11, 17, 23, 24, 25, 26, 27] | 1.4637 | -0.272 | 0.6332 | 0.7294 |
| 12 | [23, 17, 22, 24, 25, 26, 27] | 1.3457 | 1.1066 | 0.3454 | 0.8346 |
| 13 | [17, 23, 24, 25, 26, 27, 18] | 1.272 | 0.7275 | 0.4899 | 0.6011 |
| 14 | [11, 17, 23, 22, 21, 20, 19] | 1.2513 | 0.5937 | 0.3554 | 0.6022 |
| 15 | [23, 22, 24, 25, 17, 11, 10] | 1.2482 | 0.684 | 0.4452 | 0.2052 |
| 16 | [23, 22, 17, 11, 24, 25, 29] | 1.217 | 0.7244 | 0.243 | 0.4678 |
| 17 | [12, 11, 17, 23, 24, 25, 29] | 1.0815 | 0.7428 | 0.325 | 0.6769 |
| 18 | [23, 17, 22, 24, 25, 26, 29] | 0.9787 | 1.2251 | 0.4922 | 0.1252 |
| 19 | [17, 23, 24, 25, 29, 36, 37] | 0.874 | 0.6954 | 0.2604 | 0.5107 |
| 20 | [12, 11, 17, 23, 22, 21, 20] | 0.7908 | 0.6826 | 0.499 | 0.9242 |
| 21 | [17, 23, 24, 25, 29, 36, 35] | 0.7654 | 0.5903 | 0.3705 | 0.237 |
| 22 | [10, 11, 17, 23, 24, 25, 26] | 0.7585 | 0.9996 | 0.5895 | 0.5356 |
| 23 | [10, 11, 17, 23, 22, 21, 20] | 0.6785 | 1.0484 | 0.4258 | 0.4572 |
| 24 | [23, 17, 22, 24, 25, 29, 36] | 0.6368 | 0.8306 | 0.6417 | 0.4367 |
| 25 | [10, 11, 17, 23, 24, 25, 29] | 0.5716 | 0.5358 | 0.6655 | 0.3611 |
| 26 | [25, 29, 26, 24, 23, 17, 11] | -0.377 | 0.016 | 0.2381 | 0.565 |

| 27 | [25, 29, 26, 27, 24, 23, 17] | -0.439 | -0.1191 | 0.4294 | 0.8693 |
| 28 | [25, 26, 29, 36, 24, 23, 17] | -0.4842 | 0.6384 | 0.5519 | 0.7687 |
| 29 | [21, 28, 20, 22, 23, 17, 11] | -1.109 | 0.9483 | 0.4192 | 0.5905 |

**Supplement C**

Here the relation between the entangle strength and the radii in the $\langle M_2 \rangle - \langle M_2' \rangle$ plane are analytically calculated for the 2-qubit case. Two qubits states with an entangled parameter $\theta$ is defined as below:

$|\Psi\rangle = (sin\theta|00\rangle + cos\theta e^{i\varphi}|11\rangle)$  (C.1)

The entangled parameter $\theta$ represents the influence from noise environment. For $\theta = \frac{\pi}{4}$, the state gets back to the fully entangle GHZ-like state, and for $\theta = 0$ or $\frac{\pi}{2}$, the superposition totally breaks down. The expectation values are calculated with this parametrically entangled GHZ-like state, and

$$\begin{cases} \langle XX \rangle = sin2\theta cos\varphi \\ \langle XY \rangle = sin2\theta cos(\varphi - \frac{\pi}{2}) \\ \langle YY \rangle = -sin2\theta cos\varphi \end{cases} \quad (C.2)$$

To get the expectation values for $\langle M_2 \rangle$ and $\langle M_2' \rangle$, the results are:

$$\langle M_2 \rangle = 2\sqrt{2}sin2\theta cos\left(\varphi - \frac{\pi}{4}\right)$$

$$\langle M_2' \rangle = -2\sqrt{2}sin2\theta sin\left(\varphi - \frac{\pi}{4}\right) \quad (C.3)$$

The radius in the $\langle M_2 \rangle - \langle M_2' \rangle$ plane is:

$$\sqrt{\langle M_2 \rangle^2 + \langle M_2' \rangle^2} = 2\sqrt{2}sin2\theta \quad (C.4)$$

From Eq. (C.4), the entangled parameter $\theta$ directly relates to the radii of $\langle M_2 \rangle$ and $\langle M_2' \rangle$. According to the influence of noise environment on the real device, different radii of the entangled strength with different parameter $\theta$ can be obtained in measurement. For example, maximal value $2\sqrt{2}$ is at $\theta = \frac{\pi}{4}$ which is a GHZ-like state and minimal values zero is at either $\theta = 0$ or $\frac{\pi}{2}$, which is the state without superposition.

Figure 1:
The connectivity of qubits within hexagonal lattice structure of IBM Q 53-qubits quantum computer (IBM Rochester).

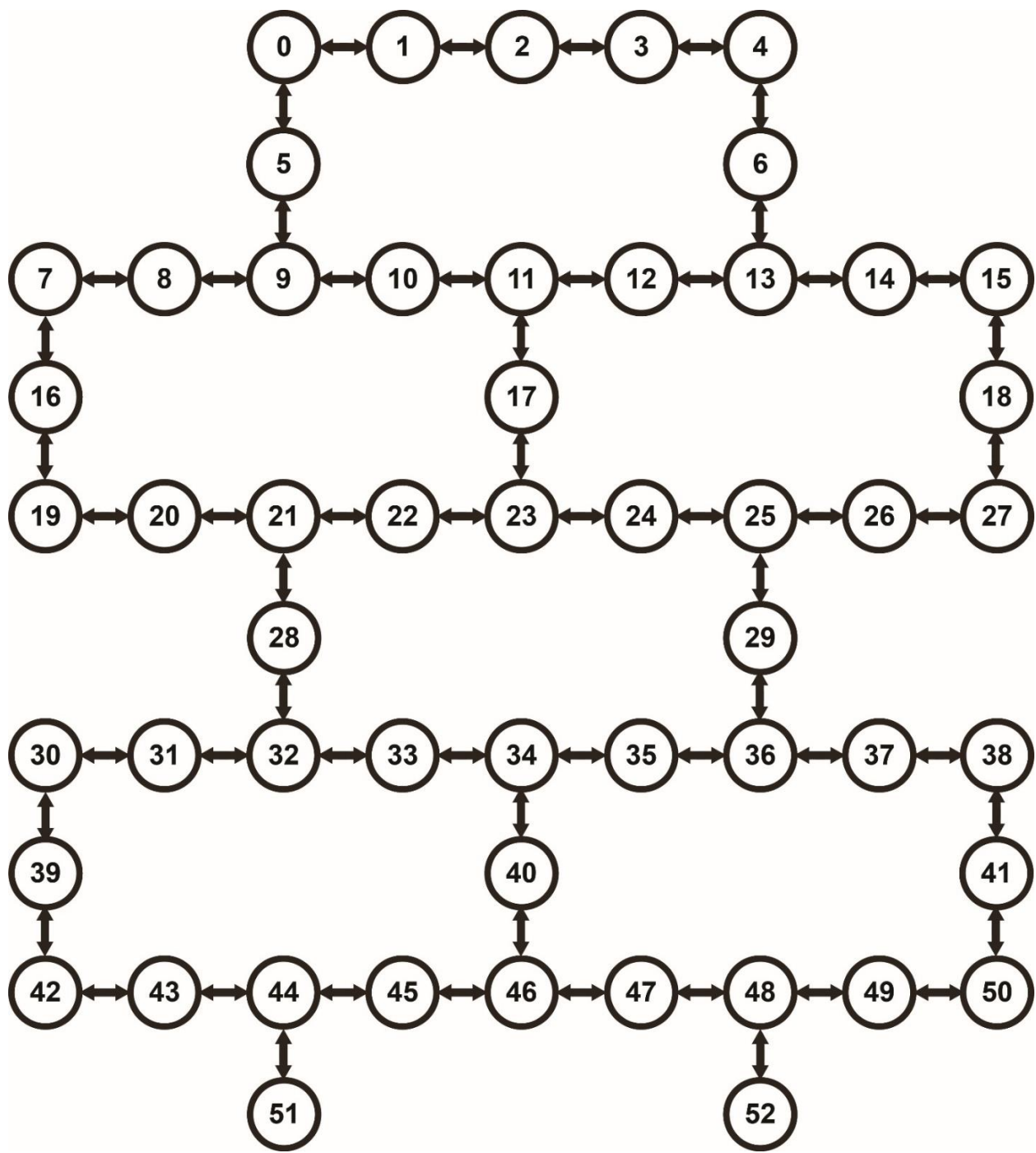

Figure 2:
(a) The quantum circuit for 5-qubits GHZ-like state, $|\text{GHZ}_5\rangle = \frac{1}{\sqrt{2}}(|00000\rangle - |11111\rangle)$. *H* gate is Hadamard gate, *U1* ($\varphi$) is a gate to rotate z with phase $\varphi$ and *CNOT* gate entangles two qubits together. (b) The quantum circuit of *XXYYY measurement for 5-qubits*.

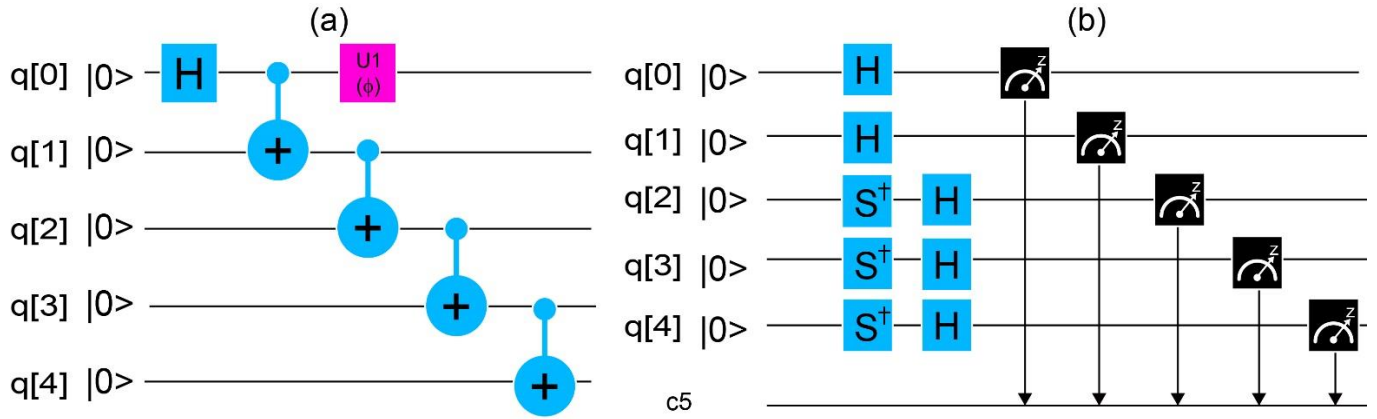

Figure 3:
The N-qubits GHZ-like states measurement of maximal expectation value of Mermin's polynomials. (a) 2 qubits (b) 3 qubits (c) 4 qubits (d) 5 qubits (e) 6 qubits (f) 7 qubits. The horizontal axis is the sequence number of connectivity for the N-qubits (listed in *supplement B*) chosen from the IBM Rochester hexagonal structure in Fig. 1. The vertical axis is $\widetilde{M_n}$, the measurement values of the Mermin's polynomials normalized by the maximal LR values. The measurement values are arranged in a descending order from left to right and the real connectivity of the qubits under measurement are listed in *Supplement B*. Experimental shots are 1024 and each data are the average of five measurements, and both the average values and standard variations (the bar length of the data point) are plotted.

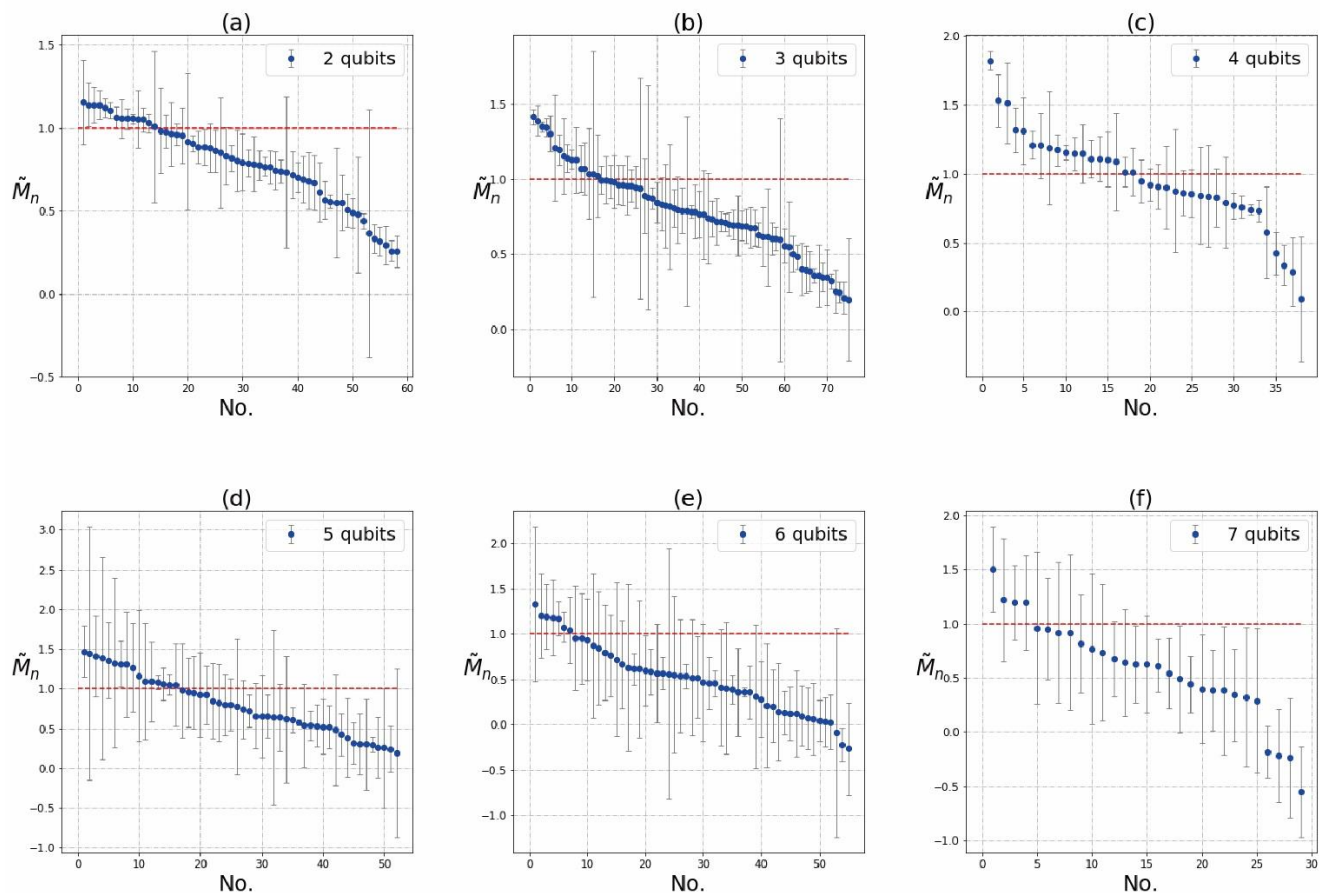

Figure 4:
The relationship between orthogonal measurements of $\widetilde{M_n}$ and $\widetilde{M'_n}$ is shown. The square corner is the upper limit for LR, the inner circle is the theoretical value of the 2-qubit subsystem entangled for a N-qubit system and outer circle is the for the 3-qubit subsystem of a N-qubit system. Experimental shots are 1024, and each data are the average of five measurements. (a) The experimental results of 2, 3, 4 qubits from different connectivity of IBM 53-qubit Rochester system. (b) The experimental results of 5, 6, 7 qubits from different connectivity of IBM 53-qubit Rochester system.

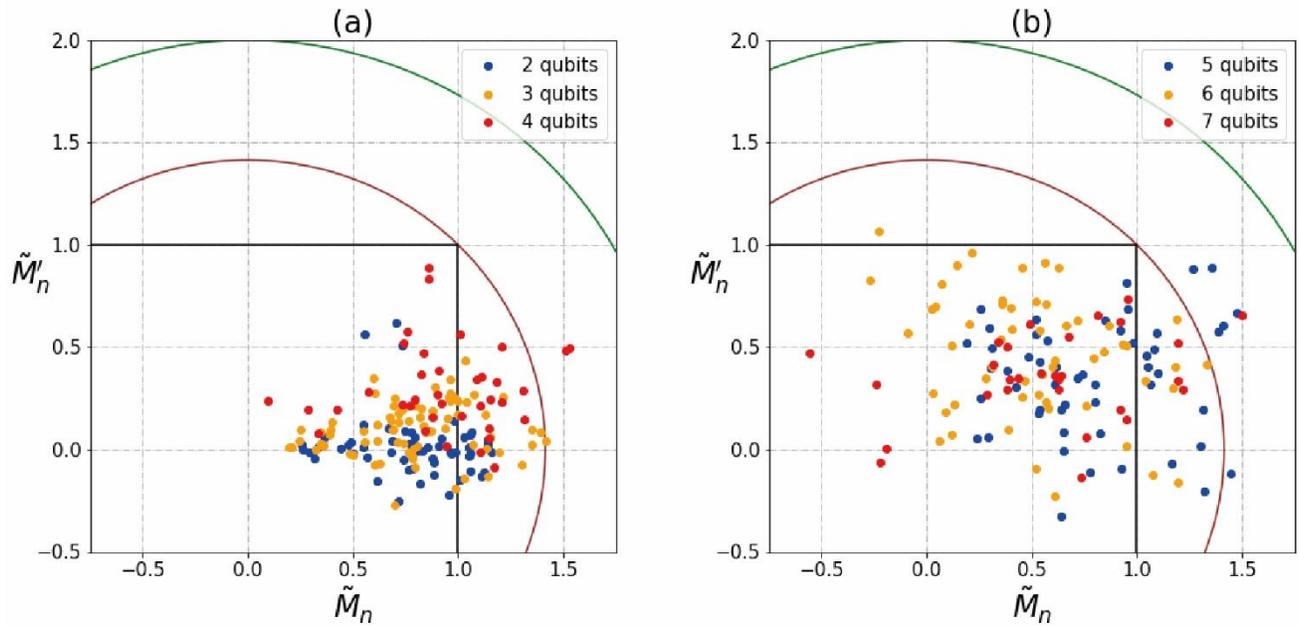

Table I: Maximal values of $\langle M_n \rangle$ and $\langle M'_n \rangle$ are normalized by $\langle M_n \rangle_{LR}$. Here $\varphi_{max}$ is phase angle which $\langle M_n \rangle$ has its maximal value.

| number of qubits: $n=$ | 2 | 3 | 4 | 5 | 6 | 7 |
|---|---|---|---|---|---|---|
| $\varphi_{max}$ | $\frac{1}{4}\pi$ | $\frac{1}{2}\pi$ | $\frac{3}{4}\pi$ | $\pi$ | $\frac{5}{4}\pi$ | $\frac{3}{2}\pi$ |
| $e^{i\varphi_{max}}$ | $\frac{1+i}{\sqrt{2}}$ | $i$ | $\frac{-1+i}{\sqrt{2}}$ | $-1$ | $\frac{-1-i}{\sqrt{2}}$ | $-i$ |
| $\frac{\text{maximal value}}{\langle M_n \rangle_{LR}} = 2^{\frac{n-1}{2}}$ | $\sqrt{2}$ | $2$ | $2\sqrt{2}$ | $4$ | $4\sqrt{2}$ | $8$ |